Oscillatory Spin Polarization and Magneto-Optic Kerr Effect in Fe$_3$O$_4$ Thin Films on GaAs(001)

Yan Li[1], Wei Han[1], A. G. Swartz[1], K. Pi[1], J. J. I. Wong[1], S. Mack[2], D. D. Awschalom[2], and R. K. Kawakami[1*]

[1]Department of Physics and Astronomy, University of California, Riverside, CA 92521

[2]Center for Spintronics and Quantum Computation, University of California, Santa Barbara, CA 93106

* email: roland.kawakami@ucr.edu

**Abstract:**

The spin dependent properties of epitaxial Fe$_3$O$_4$ thin films on GaAs(001) are studied by the ferromagnetic proximity polarization (FPP) effect and magneto-optic Kerr effect (MOKE). Both FPP and MOKE show oscillations with respect to Fe$_3$O$_4$ film thickness, and the oscillations are large enough to induce repeated sign reversals. We attribute the oscillatory behavior to spin-polarized quantum well states forming in the Fe$_3$O$_4$ film. Quantum confinement of the $t_{2g}$ states near the Fermi level provides an explanation for the similar thickness dependences of the FPP and MOKE oscillations.

PACS: 85.75.-d, 75.70.-i, 78.47.J-, 73.21.Fg

Spintronics strives to revolutionize semiconductor electronics by utilizing the electron's spin in addition to charge for integrated memory and logic functions [1]. For spin injection and detection, magnetite ($Fe_3O_4$) is an attractive material because theory predicts complete density of states (DOS) spin polarization at the Fermi level (i.e. half metal) [2, 3], experiments measure a large spin polarization (55% - 80%) [4, 5], and the Curie temperature of 858 K is much higher than room temperature. Following the successful growth of $Fe_3O_4$ thin films on GaAs [6], the recent demonstration of spin injection establishes $Fe_3O_4$ as an important material for semiconductor spintronics [7].

One interesting aspect of magnetic thin films and multilayers is the confinement of electron waves to form quantum well (QW) states. Because the QW states are spin-polarized, this produces oscillatory interlayer magnetic coupling [8, 9] and modulates magnetic properties such as the magnetic anisotropy [10] and magneto-optic Kerr effect (MOKE) [11, 12]. $Fe_3O_4$ is particularly appealing in this regard because its relatively low carrier density ($\sim 10^{21}$ cm$^{-3}$) compared to metals and large spin polarization should lead to strong modulation of spin dependent properties that could be tuned by electrostatic gates. While evidence for quantum confinement in $Fe_3O_4$ has been reported for thin films and nanoparticles [13, 14], their effect on spin dependent properties has not been established.

In this Letter, we report strong oscillations and sign reversals in the spin polarization and MOKE of $Fe_3O_4$ films as a function of thickness, which we attribute to the formation of spin-polarized QW states. High quality $Fe_3O_4$ films on GaAs(001) are fabricated by molecular beam epitaxy (MBE), and the Fermi level spin polarization of $Fe_3O_4$ is probed using the ultrafast optical measurement of ferromagnetic proximity polarization (FPP) [15, 16]. The systematic thickness dependence of FPP and MOKE are measured on wedged $Fe_3O_4$ films on GaAs(001), and similar oscillatory behaviors are observed even though the two measurements rely on different mechanisms (spin dependent electron reflection for FPP, optical transitions for MOKE). Quantum confinement of the $t_{2g}$ states near the Fermi level provides an explanation for the similar thickness dependences of the FPP and MOKE oscillations. Our results demonstrate the tuning of spin dependent properties of $Fe_3O_4$/GaAs hybrid structures by quantum confinement, suggesting potential applications in semiconductor spintronic devices.

Samples are grown by MBE with the following structure: Al cap(2 nm)/$Fe_3O_4$/GaAs(123 nm)/$Al_{0.7}Ga_{0.3}As$(400 nm)/GaAs(001), with n-type doping of the GaAs epilayer (Si: $7\times10^{16}$ cm$^{-3}$). The GaAs template is grown in a separate III-V chamber and capped with As, transferred in air to a second chamber for $Fe_3O_4$ growth, and the As is desorbed to produce a (2×4) surface reconstruction. A single crystalline Fe (5 nm) film is deposited at RT, with thickness determined by a quartz deposition monitor. Next, molecular oxygen is leaked into the vacuum chamber ($P_{O2}$ = $5\times10^{-7}$ torr [6, 17]) and the sample is heated to 175 °C for the formation of $Fe_3O_4$. The RHEED pattern evolves from Fe streaks into a typical $Fe_3O_4$ RHEED pattern [6, 7] within 3 minutes after reaching 175 °C. Figures 1(a) and 1(b) show the corresponding RHEED patterns along the [110] and [010] in-plane directions of GaAs after 30 min of oxidation, indicating epitaxial growth of $Fe_3O_4$ (10 nm) on GaAs. We assume the Fe is completely oxidized and 1 nm of Fe corresponds to 2.086 nm of $Fe_3O_4$.

The magnetic properties of the $Fe_3O_4$ films are characterized by MOKE and vibrating sample magnetometry (VSM). Figure 1(c) shows the hysteresis loop of $Fe_3O_4$ (10 nm) measured at 80 K along the [110] in-plane direction of GaAs, and other directions show almost identical magnetic behavior. The magnetization measured by VSM at RT is 4.1±0.1 Bohr magnetons ($\mu_B$) per $Fe_3O_4$ formula unit (Figure 1c inset). The square hysteresis loop with magnetization value close to the ideal value of $4\mu_B/Fe_3O_4$ [2, 3] indicates the high quality of the film, large magnetic domains, and the absence of antiphase boundaries [18]. Figure 1(d) shows the temperature dependence of the remanent magnetization as determined by the MOKE measurement. Upon cooling from RT to 4 K, the magnetization increases slightly as typical of ferrimagnetic behavior, and the coercivity also increases as the temperature is lowered. There is a small decrease in magnetization as the sample is cooled below 120 K, which is a suppressed Verwey transition, consistent with previous studies of ultrathin $Fe_3O_4$ films [19, 20].

We investigate the spin polarization of the $Fe_3O_4$ film through time-resolved Faraday rotation (TRFR) measurements of the FPP effect. This is described in detail in the supplementary online materials [21] and in ref [22]. Briefly, unpolarized conduction electrons are optically excited in the GaAs by a linearly polarized pump pulse tuned near the band gap. Immediately following the excitation (within 50 ps [16]), the electrons gain in-plane spin polarization, $S_{FPP}$, by reflecting off the FM/GaAs interface (i.e. FPP effect). We are most interested in the value of $S_{FPP}$ because it provides a measure of the spin polarization of the FM's density of states at the Fermi level [23, 24]. Because the Faraday rotation of a linearly-polarized, normally incident probe beam is proportional to the out-of-plane component of spin while the $S_{FPP}$ is oriented in-plane, $S_{FPP}$ is determined by measuring electron spin dynamics in a tilted magnetic field ($B_{app}$, oriented 30° out of plane) which generates the measurable out-of-plane component, as illustrated in the Figure 2 insets. The out-of-plane component of the Larmor spin precession about a cone is given by:

$$S_z = \frac{\sqrt{3}}{4} S_{FPP} (\hat{m} \cdot \hat{e}_y) \left( \exp(-\Delta t / T_1) - \cos(2\pi\nu_L \Delta t) \exp(-\Delta t / T_2^*) \right) \qquad (1)$$

where $\nu_L = g\mu_B B_{app} / h$ is the Larmor frequency, $g$ is the g-factor of GaAs (-0.44), $\mu_B$ is the Bohr magneton, $\Delta t$ is the pump-probe time delay, $\hat{m}$ is the unit vector along the magnetization, and the $T_1$ and $T_2^*$ are longitudinal and transverse spin lifetimes, respectively. For convenience, we express $S_z$ and $S_{FPP}$ in units of the Faraday rotation angle. The value of $S_{FPP}$ is extracted by fitting the TRFR data with this equation.

Figures 2(a) and 2(b) show TRFR curves on Fe(4 nm)/GaAs and Fe$_3$O$_4$(8 nm)/GaAs prepared by oxidation at 175 °C for 60 min. The delay scan for the Fe(4 nm)/GaAs hybrid structures show positive $S_{FPP}$, consistent with previous studies [16, 22]. On the other hand, the $S_{FPP}$ for Fe$_3$O$_4$/GaAs is negative ($S_{FPP}$ = -73 μrad) and has a larger magnitude than for the Fe/GaAs sample ($S_{FPP}$ = 28 μrad). The opposite $S_{FPP}$ in Fe$_3$O$_4$/GaAs compared to Fe/GaAs is expected because the DOS at the Fermi level in bcc Fe has a positive spin polarization (majority spin) while the Fermi level DOS of Fe$_3$O$_4$ is theoretically predicted to have 100% negative spin polarization [2, 3]. This result demonstrates the spin dependent reflection in Fe$_3$O$_4$/GaAs and the sign and magnitude of $S_{FPP}$ are consistent with theoretical expectations.

We utilize wedged Fe$_3$O$_4$ films to investigate the thickness dependence of the FPP. Figure 3 shows $S_{FPP}$ vs. film thickness for four different oxidation temperatures (150 °C, 175 °C, 225 °C, and 275 °C). For all samples, the RHEED patterns are characteristic of Fe$_3$O$_4$ (see supplementary material, Figure S3) [21]. Interestingly, the curves for 150 °C and 175 °C exhibit oscillations in $S_{FPP}$ as a function of film thickness. For the 150 °C sample, $S_{FPP}$ oscillates between negative and positive values through nearly two oscillations with a period of ~4.2 nm. For the 175 °C sample, $S_{FPP}$ oscillates mainly between a negative value and zero with a period of ~5.0 nm. The oscillatory behaviors have been observed on two samples prepared at 150 °C and two samples prepared at 175 °C, with consistent results for sign reversal and period (within 20%). For oxidation at 225 °C, $S_{FPP}$ is negative and the oscillations are no longer present. At the higher oxidation temperature of 275 °C, no FPP signal is observed.

The oscillations in $S_{FPP}$ as a function of thickness could be explained by the formation of spin-polarized QW states in the $Fe_3O_4$ film, which causes the Fermi level spin polarization of the $Fe_3O_4$ to oscillate between positive and negative values. In principle, there are two distinct wavelengths for spin up and spin down electrons in magnetic films and QW states form according to the quantization condition

$$2k_{\uparrow,\downarrow}d + \phi_{\uparrow,\downarrow} = 2\pi n \qquad (2)$$

where $n$ is an integer, $d$ is the film thickness, and $k_\uparrow$ ($k_\downarrow$) and $\phi_\uparrow$ ($\phi_\downarrow$) are the wavevector and the phase accumulated for spin up (down) electrons upon reflection at the boundaries, respectively. This produces oscillations in the Fermi level ($\varepsilon_F$) DOS as a function of thickness with periods of $\pi/k_\uparrow(\varepsilon_F)$ and $\pi/k_\downarrow(\varepsilon_F)$. Theoretically, $Fe_3O_4$ is predicted to be a half-metal with only spin down electrons, but spin-polarized photoemission experiments find DOS spin polarization below 80% [4, 5]. Therefore, we consider both spin polarizations at the Fermi level. For the 150 °C sample, the sign of $S_{FPP}$ oscillates between positive and negative values, indicating that both spin states are present at the Fermi level. The oscillation could be due to quantum confinement of one spin species or both spin species. For the 175 °C sample, $S_{FPP}$ oscillates between zero and negative values, so it is possible that there are only spin down states confined at $\varepsilon_F$. The oscillation period of 4-5 nm is longer than typical periods observed in metallic QWs (less than 1 nm), which could result from smaller Fermi wavevectors associated with the lower electron density of $Fe_3O_4$ ($\sim 10^{21}$ cm$^{-3}$ [25, 26]) compared to metals ($\sim 10^{23}$ cm$^{-3}$). Furthermore, earlier work on QW states in $Fe_3O_4$ [13, 14], suggests a de Broglie wavelength ($2\pi/k_F$) of ~10 nm (i.e. QW oscillation period of ~5 nm), which is consistent with our data. A quantitative investigation of the oscillatory period would require a direct comparison of the Fermi surface and QW states via angle-resolved photoemission spectroscopy (ARPES) [27]. At higher temperatures, the interface is expected to degrade through interdiffusion or over-oxidation and therefore the QW states would be destroyed. Apart from the spin polarization of the Fermi level DOS of the $Fe_3O_4$ film, there are other factors that could affect the sign of the FPP signal such as the Schottky barrier height and carrier concentration of the GaAs [23, 24]. However, these effects are ruled out as the origin of the FPP oscillations as discussed below.

To further explore the origin of the oscillations, we measure the magnetic properties of the $Fe_3O_4$/GaAs hybrid structure along the wedge by longitudinal MOKE (835 nm at RT and 80 K), which depends only on the properties of $Fe_3O_4$ layer. Figure 4(a) - (c) shows the MOKE hysteresis loops taken on the 150 °C sample at different film thicknesses at 80 K. Interestingly, the sign of the MOKE also depends on the film thickness. A more detailed scan of the remanent MOKE signal as a function of thickness (Figure 4d) displays oscillations at 80 K and 300 K with similar shape as the FPP thickness dependence. This implies that the oscillations in both FPP and MOKE are related to the properties of $Fe_3O_4$ layer, as opposed to Schottky barrier or parameters of GaAs. While the quantitative calculation of the MOKE coefficient in magnetic metals is rather complicated [28-30], it is known that the photoexcitation process is determined by Fermi's golden rule, with $h\nu = \varepsilon_f(k) - \varepsilon_i(k)$, where $\varepsilon_i(k)$ and $\varepsilon_f(k)$ are initial and final states of photoexcitation, respectively. Modulating the density of the initial or final states by quantum confinement should induce oscillations in the MOKE. Theoretically, the Kerr rotation in $Fe_3O_4$ for 1.49 eV photon energy (835 nm) involves the minority $t_{2g}$ states [29, 30]. Because the $t_{2g}$ states are also responsible for the spin polarization at the Fermi level [2, 3], quantum confinement of the $t_{2g}$ bands can account for the oscillations in both the MOKE and FPP signals. Moreover, this explains why two experiments that rely on different physical processes (optical transitions for MOKE, electron reflection for FPP) exhibit similar oscillatory behavior. The differences in the oscillation periods of MOKE and FPP are likely due to the fact that the MOKE not only depends on the states at the Fermi level but also states away from the Fermi level, which can produce different periods because of the energy dispersion of the wavevector [26].

In conclusion, we have successfully fabricated epitaxial $Fe_3O_4$ films on GaAs(001) by post oxidation of single crystalline Fe and observe oscillations in both the FPP and MOKE signals as a function of film thickness. The oscillations are strong enough to induce sign reversals of both the spin polarization via FPP and the Kerr rotation. We attribute the oscillatory behavior to the formation of spin-polarized QW states in the $Fe_3O_4$ film. Future studies utilizing direct probes of the electronic structure (e.g. ARPES [27]) should be performed to further investigate the oscillatory behavior.

We acknowledge technical assistance from T. Lin and J. Shi. This work was supported by NSF, CNN/DMEA, and ONR.

**FIGURE CAPTIONS:**

**Figure 1.** (a, b) RHEED patterns of $Fe_3O_4$(10 nm)/GaAs(001) along [110] and [010] directions of GaAs. The beam energy is 15.0 keV. (c) Hysteresis loop measured by longitudinal MOKE at 80 K. Inset: hysteresis loop measured by VSM at RT. (d) Remanent magnetization vs. temperature measured by MOKE.

**Figure 2.** (a, b) Representative TRFR curves on Fe(4 nm)/GaAs and $Fe_3O_4$(8 nm)/GaAs, respectively. The measurements are performed at 80 K and with a linearly polarized pump pulse to directly measure the FPP. The open squares are data points and the solid lines are fits by Eq. (1). Insets: FPP measurement geometry.

**Figure 3.** $S_{FPP}$ as a function of thickness for wedge samples oxidized at 150 °C, 175 °C, 225 °C and 275 °C, respectively. Error bars for all the data points are displayed, or smaller than the size of the data points.

**Figure 4.** (a) – (c) MOKE hysteresis loops on $Fe_3O_4$ thicknesses at 3.2 nm, 4.8 nm and 7.6 nm, respectively. (d) Remanent MOKE (Kerr rotation) along the wedge sample oxidized at 150 °C, measured at 80 K (black squares) and 300 K (black triangles, offset by -10 mrad), and $S_{FPP}$ (red/grey square) is plotted for comparison. The errors of the data points are within the size of the data points.

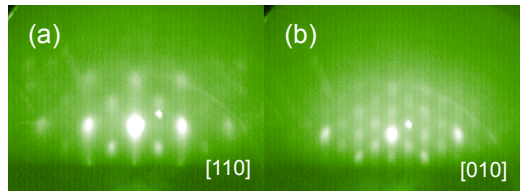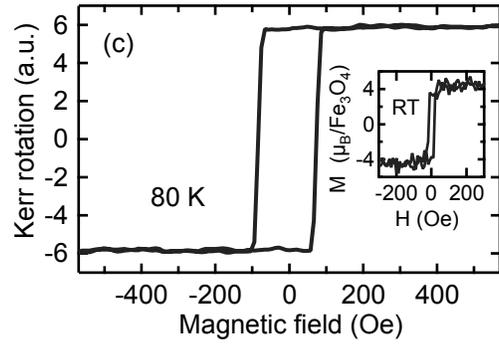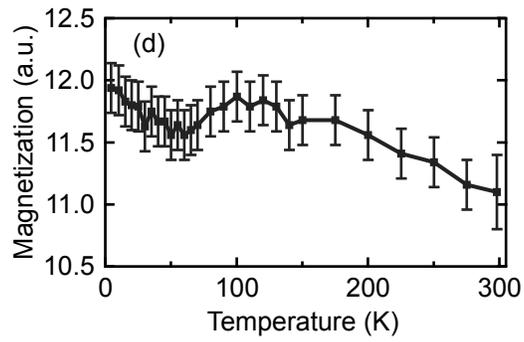

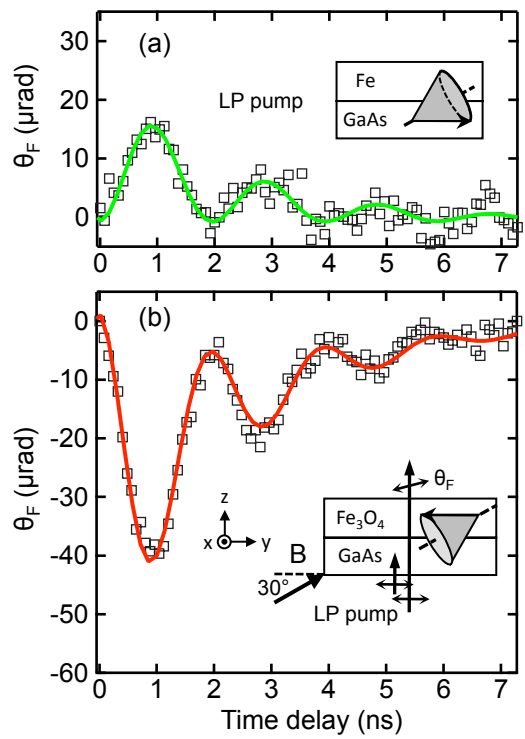

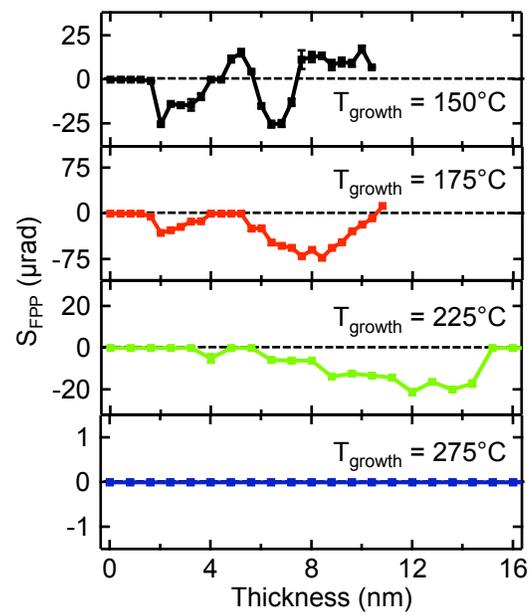

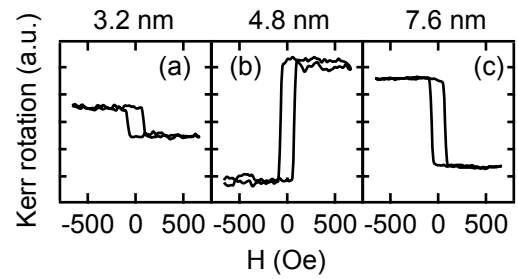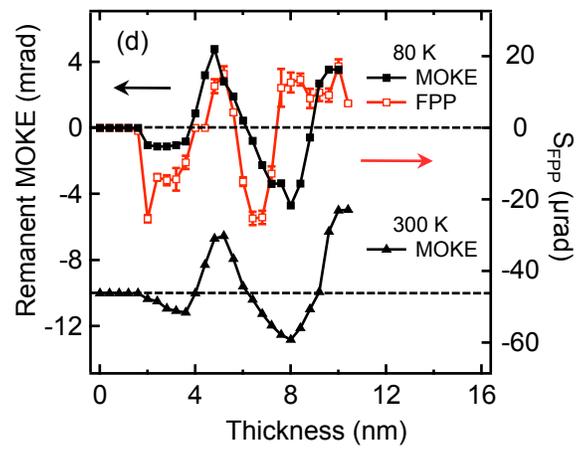

Oscillatory Spin Polarization and Magneto-Optic Kerr Effect in $Fe_3O_4$ Thin Films on GaAs(001)

Yan Li[1], Wei Han[1], A. G. Swartz[1], K. Pi[1], J. J. I. Wong[1], S. Mack[2], D. D. Awschalom[2], and R. K. Kawakami[1*]

[1]Department of Physics and Astronomy, University of California, Riverside, CA 92521

[2]Center for Spintronics and Quantum Computation, University of California, Santa Barbara, CA 93106

**Supplementary Material**

1. Ferromagnetic Proximity Polarization Measurement
2. RHEED patterns of $Fe_3O_4$ films on GaAs(001) prepared at different temperatures

## 1. Ferromagnetic Proximity Polarization Measurement

Overview:

In this work, we utilize the ultrafast optical measurement of ferromagnetic proximity polarization (FPP) to probe the spin polarization of $Fe_3O_4$ films. The technique is conceptually similar to spin-polarized electron microscopies (e.g. spin-polarized low energy electron microscopy, scanning electron microscopy with polarization analysis, etc.) where electron beams probe the spin polarization of vacuum/ferromagnet (FM) interfaces. Figure S1a shows a scattering process which gives rise to a spin polarized beam due to spin dependent reflection of an unpolarized beam from a ferromagnetic surface. By measuring the electron spin of the scattered beam, one can infer the magnetic properties and spin-polarization of the FM surface. As shown in Figure S1b, FPP is the solid state analog, where the electrons are inside a semiconductor (GaAs) instead of vacuum. The unpolarized incident electrons are generated in GaAs by a linearly-polarized optical pump pulse, and the polarization of the reflected electrons is detected by the Faraday rotation of a linearly-polarized optical probe pulse. The FPP occurs within the first 50 ps after the optical excitation [1]. Measuring the spin dynamics of the reflected electrons (time after 50 ps) by time-resolved Faraday rotation (TRFR) reveals the initial electron spin and thus the spin polarization of the FM film at the Fermi level [2], which is the quantity of the most interest in our study of the $Fe_3O_4$ films.

The FPP measurement is different from typical TRFR measurements that employ a circularly-polarized pump pulse to generate spin polarized electrons in the GaAs [3]. In those studies, the focus is usually on the spin dependent properties of the GaAs such as the electron spin dynamics, spin relaxation in GaAs, spin manipulation, etc. In the present study, the TRFR measurement of electron spin dynamics in GaAs is utilized as a tool to investigate the Fermi level spin polarization of the $Fe_3O_4$ films. The measurement is complicated by the fact that the Faraday rotation is sensitive to the out-of-plane component of the electron spin, but the reflected electrons have spin oriented in-plane. Thus, a canted magnetic field (Figure S2a) is applied to induce precession of the electron spin, which generates an out-of-plane spin component for detection by the Faraday effect, as shown in Figure S2b.

Experimental Details:

For the optical transmission measurement, the sample is mounted onto sapphire and the GaAs substrate is removed by a selective chemical etch ($NH_4OH/H_2O_2$ spray etch) that stops at the $Al_{0.3}Ga_{0.7}As$ layer. The measurement is performed at $T = 80$ K which provides sufficient spin lifetime while avoiding the effects of ferromagnetic imprinting of nuclear spins [4] and resonant spin amplification [3]. Laser pulses (150 fs, 76 MHz repetition rate) tuned near the band gap of GaAs ($\lambda = 812$ nm) are generated by a Ti:sapphire laser and are focused to a ~40 μm diameter spot on the sample. The average powers for the pump and probe beams are 2 mW and 0.05 mW, respectively. The pump and probe beams have normal incidence (z-axis), and the applied magnetic field $B_{app}$ is tilted out of the film plane by an angle $\alpha = 30°$ (in the y-z plane), as shown in Figure S2a. $B_{app}$ is set to 900 G and we assume that the magnetization remains in the film plane (x-y) due to magnetic shape anisotropy. With a linearly polarized pump beam, unpolarized electrons are excited in the GaAs layer (Figure S1b). Due to the spin dependent DOS and the associated wavevectors of the FM at the Fermi level, the electron wave experiences a spin dependent transmission and reflection at the FM/GaAs interface. This generates a spontaneous spin polarization $S_{FPP}$ in the GaAs layer, oriented along the magnetization axis of the ferromagnet layer (Figure S1b), which occurs within 50 ps of the optical excitation. We adopt a sign convention that $S_{FPP}$ is positive (negative) when the initial spin is parallel (antiparallel) to the magnetization $M$. The $B_{app}$ induces spin precession about a cone (Figure S2a) to generate a z-component of spin, and the dynamics is measured by a time-delayed, linearly-polarized probe pulse via Faraday rotation ($\theta_F \propto S_z$) of the polarization axis (Figure S2a). Quantitatively, the spin dynamics for this population includes both a transverse component (with lifetime $T_2^*$) and a longitudinal component (with lifetime $T_1$), and the TRFR signal is given by:

$$S_z = S_{FPP}\left(\hat{m} \cdot \hat{e}_y\right)\sin\alpha\cos\alpha\left[\exp(-\Delta t / T_1) - \cos(2\pi v_L \Delta t)\exp(-\Delta t / T_2^*)\right] \quad (S1)$$

where $v_L = g\mu_B B / h$ is the Larmor precession frequency, $g$ is the g-factor of GaAs ($g \approx -0.44$), $\mu_B$ is the Bohr magneton, $h$ is Planck's constant, $\Delta t$ is the pump-probe time delay, and $\hat{m}$ is the unit vector of the magnetization. For this equation, the interaction time of the FPP (< 50 ps) is neglected, which is reasonable because the spin dynamics is in the time scale of ns. Because the transformation ($B_{app} \rightarrow -B_{app}$, $\mathbf{M} \rightarrow -\mathbf{M}$) yields $S_z(\Delta t) \rightarrow -S_z(\Delta t)$, we perform time delay scans for positive and negative $B_{app}$ and subtract the two curves to eliminate background signals unrelated to FPP. A typical time delay scan of the z-component electron spin is shown in Figure S2b. At zero time, the electron spin is in-plane, and the $\theta_F \propto S_z$ is zero; as it precesses and gradually goes out-of-plane, the $\theta_F$ increases to maximum; with further precession, the spin turns in-plane again, corresponding to a minimum in the time delay scan curve. The sign and magnitude of the spin precession signal is directly related to the value of $S_{FPP}$, which is of primary interest in our measurement. Quantitatively, the value of $S_{FPP}$ is obtained by fitting the TRFR data with equation S1.

The possibility that the TRFR signal originates from precession of the FM magnetization is excluded by the following tests. First, we measure the wavelength dependence of the TRFR signal. A signal from electron spin in GaAs will reduce quickly as the wavelength is tuned away from the band gap, while a signal from the FM magnetization will not. Second, we measure the precession frequency as a function of magnetic field. For electron spin precession in GaAs, the frequency increases linearly with the field and the proportionality constant should yield |g| = 0.44. For magnetization precession, the frequency is determined by a combination of the applied magnetic field and anisotropy field, which is not linear with the applied field. Third, we lower the temperature down to 5 K to verify the FPP phenomena by observing the resulting ferromagnetic imprinting of nuclear spins [4]. We have performed all these tests to verify that the TRFR signal originates from electron spin precession in the GaAs, as opposed to magnetization precession of the FM.

## 2. RHEED patterns of Fe$_3$O$_4$ films on GaAs(001) prepared at different temperatures

The FPP effect in Fe$_3$O$_4$/GaAs hybrid structures is highly sensitive to the oxidation temperature (Figure 3). The 175 °C sample shows the maximum $S_{FPP}$, and $S_{FPP}$ decreases as the temperature increases and completely goes away at 275 °C. Further structural characterization of the Fe$_3$O$_4$ films provides more insight into the degradation of FPP effect with increasing temperature. Figure S3 shows the RHEED patterns of the Fe$_3$O$_4$ (10 nm) films prepared at different oxidation temperatures, and the patterns all indicate the formation of Fe$_3$O$_4$ films with slightly sharper patterns at higher temperatures. This indicates good surface structure throughout the entire temperature range. Therefore, the decrease of the $S_{FPP}$ with increasing temperature is most likely due to degradation of the Fe$_3$O$_4$/GaAs interface.

# Figure S1

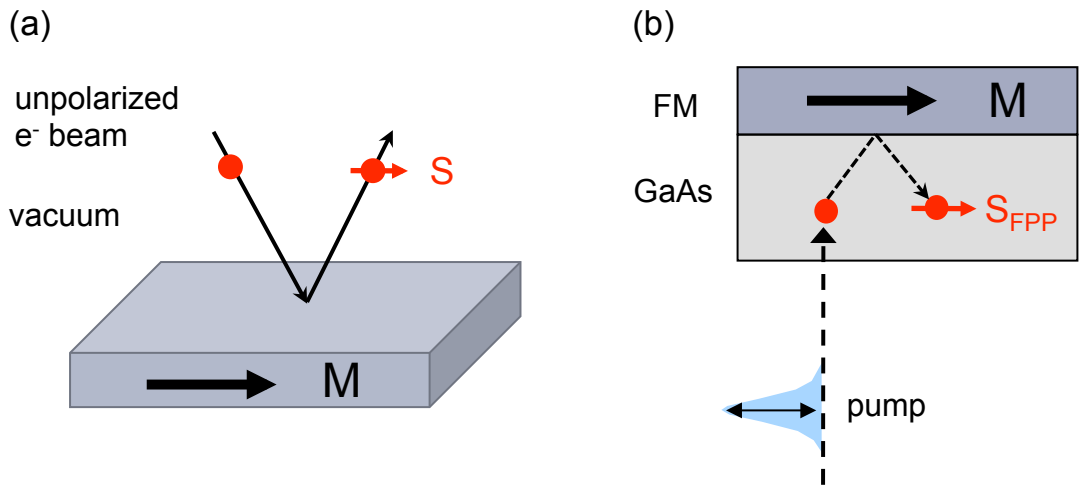

FIG S1  (a) Polarized electron reflection from magnetic film surface. (b) Ferromagnetic proximity polarization effect. Unpolarized electron carriers are excited by optical linearly polarized pump pulse. Spontaneous electron spin is generated due to spin dependent transmission and reflection at the ferromagnet/GaAs interface. The electron spin aligns along the magnetization axis of the ferromagnet layer.

Figure S2

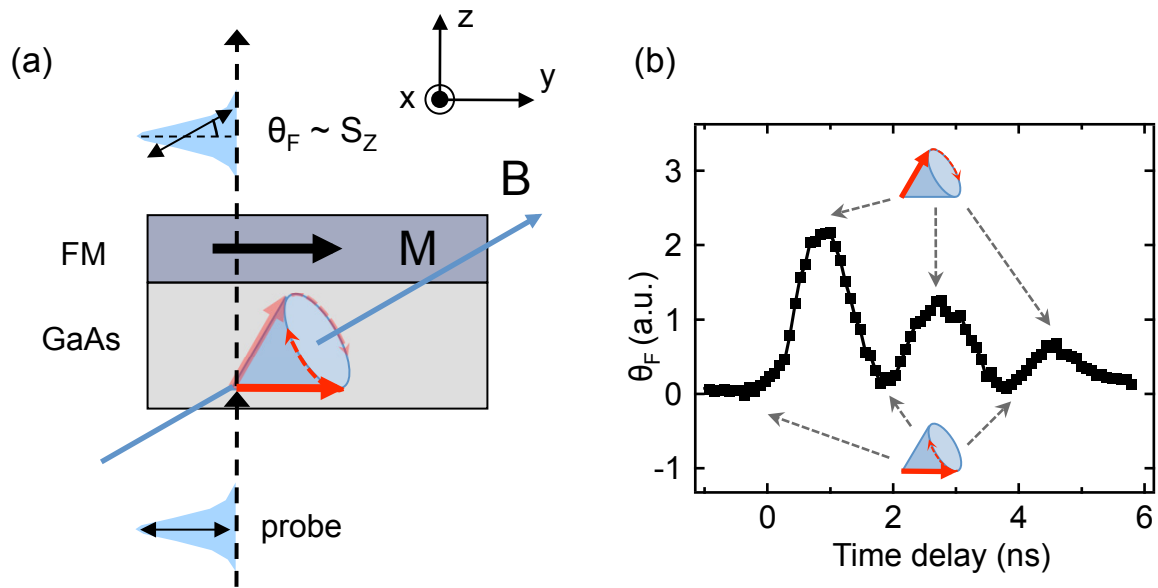

FIG S2 (a) Ultrafast optics measurement of ferromagnetic proximity polarization. Electron spins precess in a cone about the magnetic field. The magnetic field is tilted 30° away from the sample plane to produce an out-of-plane component. A time delayed probe beam pulse measures spin polarization along z-direction by Faraday rotation $\theta_F$. By varying the time delay, a curve of $\theta_F$ vs. time delay is recorded. (b) A typical time delay scan. By monitoring the spin dynamics, the initial spin polarization $S_{FPP}$ is determined.

# Figure S3

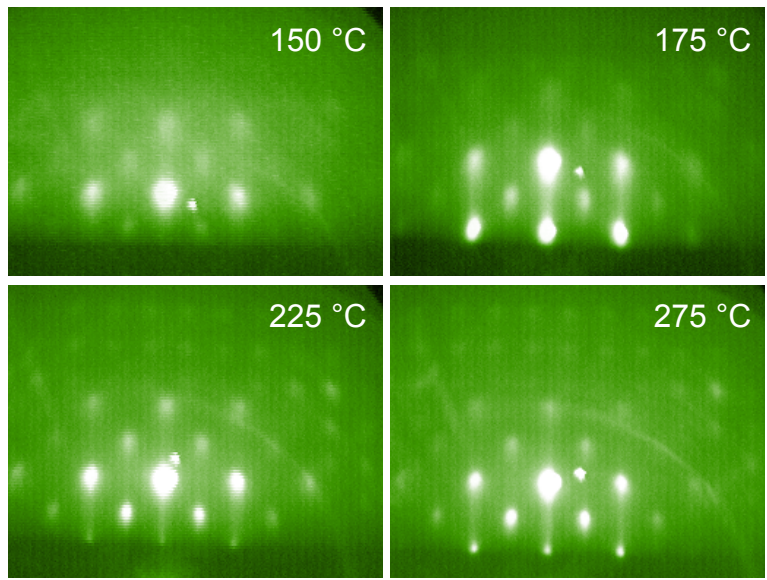

FIG. S3. RHEED patterns of Fe$_3$O$_4$(10 nm)/GaAs(001) oxidized at 150 °C, 175 °C, 225 °C and 275 °C, respectively. The diffraction is along [110] of GaAs and the beam energy is 15.0 keV. The patterns are measured at room temperature.